\documentclass[twocolumn,aps,prb,raggedbottom,nobalancelastpage,amssymb,superscriptaddress,showpacs]{revtex4-1}

\usepackage{amsmath}
\usepackage{amssymb}
\usepackage{amsfonts}
\usepackage{dsfont}
\usepackage{graphicx}
\usepackage{bm}
\usepackage{color}
\usepackage{appendix}
\usepackage{epsfig}
\usepackage{dsfont}

\newcommand{\ave}[1]{\left\langle #1\right\rangle}

\newcommand{\bra}[1]{\left<{#1}\right|}
\newcommand{\ket}[1]{\left|{#1}\right>}

\begin{document}
\title{Many-body manifestation of interaction-free measurement: the Elitzur-Vaidman bomb}
\author{Oded Zilberberg}
\affiliation{Theoretische Physik, Wolfgang-Pauli-Strasse 27, ETH Zurich, CH-8093 Zurich, Switzerland.} %
\author{Alessandro Romito}
\affiliation{\mbox{Dahlem Center for Complex Quantum Systems and Fachbereich Physik, Freie Universit\"at Berlin, 14195 Berlin, Germany}}
\author{Yuval Gefen}
\affiliation{Department of Condensed Matter Physics, Weizmann Institute of Science, Rehovot, Israel.} %

\date{\today}
\begin{abstract}
We consider an implementation of the Elitzur-Vaidman bomb experiment in a DC-biased electronic Mach-Zehnder interferometer with a leakage port on one of its arms playing the role of a ``lousy bom''. Many-body correlations tend to screen out manifestations of interaction-free measurement. Analyzing the correlations between the current at the interformeter's drains and at the leakage port, we identify the limit where the originally proposed  single-particle effect is recovered. Specifically, we find that in the regime of sufficiently diluted injected electron beam and short measurement times, effects of quantum mechanical wave-particle duality emerge in the cross-current correlations.
\end{abstract}

\maketitle

\section{Introduction}

Since its introduction, quantum mechanics has kindled the imagination
of scholars due to the interplay of its non-local character and
particle-wave duality. Using recent advances in technological control
over coherent systems, demonstration of these treats are still at the forefront of contemporary research
~\cite{haroche_nobel_2013}.  In other words, a measurement of a quantum particle (the latter may be described as a wave packet) unveils its discrete nature, when it collapses to reside at a single point. The same particle, before ``collapsing'', had assumed a non-local character. The compatibility of particle collapsing at a point and non-locality  has been discussed and demonstrated in the context of the so-called Elitzur-Vaidman (EV) bomb [aka  “interaction free measurement” (IFM)]: the wave-like interference of a single quantum particle is modified by the onset  of a measurement (bomb) performed at one of an interferometer’s arms, which could (but may not)  destroy the particle~\cite{Elitzur:1993}. 

The interferometer at hand  is tuned such that when the ``bomb''
is absent, wave-like  destructive interference renders one of its output ports dark.
One then introduces the bomb (hidden in a black box) in one of the interferometer’s arms. The bomb being ``lousy'' implies that even when a particle goes through that arm, there is a finite probability (possibly close to 1) that it will not explode. 
If the bomb eventually explodes, one knows \textit{a posteriori} that the bomb was there. But 
there is a probability that the bomb does not go off, yet  one detects a  particle
at the  interferometer's dark port. That would definitely indicate
that the black box has modified the interference pattern, hence  a bomb has been introduced inside the black box. The detection of the presence of the bomb occurs  when
no interaction with it took place. Notably, there is another possible {inconclusive}
outcome: the bomb does not go off, and the interfering particle exits at the bright port. In that case one does not know whether the bomb was there or not.
No matter how lousy the bomb is, within the many-body context of 
quantum  physics, as the signal in the interferometer
is collected over an ensemble of injected particles, there is a vanishing probability that the bomb would remain unexploded at asymptotically long times.
Rather than a bomb, the realization of this EV experimental setup requires the construction of an interferometer with an absorber
positioned on one of the interfering paths,  as well as, the introduction of a single-particle source~\cite{Kwiat:1995,Voorthuysen:1996,Hafner:1997,Tsegaye:1998,Kwiat:1999,Jang:1999,Hosten:2006,Wolfgramm:2011}. As such, this topic has remained mostly in the realm of quantum optics where IFM experiments have been proposed and demonstrated in various systems~\cite{Kwiat:1995,Voorthuysen:1996,Hafner:1997,Tsegaye:1998,Kwiat:1999,Jang:1999,Hosten:2006,Wolfgramm:2011} with a variety of applications including imaging~\cite{Kwiat:1998}, quantum computing~\cite{Hosten:2006,Vaidman:2007}, and single-photon generation~\cite{Wolfgramm:2011}. 

Interestingly, several theoretical studies of the realization and
utilization of IFM in electronic solid-state devices were recently
pursued by considering, for example, superconducting quantum-bits (qubits)~\cite{Paraoanu:2006}. Additionally, an earlier study of  electronic Mach Zehnder interferometers
(e-MZI)~\cite{Strambini:2010,Chirolli:2010}, has  focused on the employment of a wave-like picture, and the influence on the interference signal of a local perturbation in the interferometer. As such, the  particle facet of the EV picture was missing. 
Indeed, e-MZI are realized using chiral edge modes of
quantum Hall bars~\cite{Halperin1982,Wen1990}, which are 1D channels well described as collective many-body plasmonic 
waves~\cite{vonDelft:1998,PhysRevB.62.7454,ji2003electronic}. Typically,
these devices are operated at constant voltage bias leading to the
injection of numerous electrons that would eventually, with certainty, trigger the EV-bomb. We note, additionally, that single-particle excitations on top of the electron sea in quantum Hall edges have recently been obtained~\cite{Dubois:2013}. All this implies that the topic of non-locality along with wave-particle duality in complex many-electron systems is amenable to experimental studies. 

\begin{figure}
\begin{center}
\includegraphics[width=0.8\columnwidth]{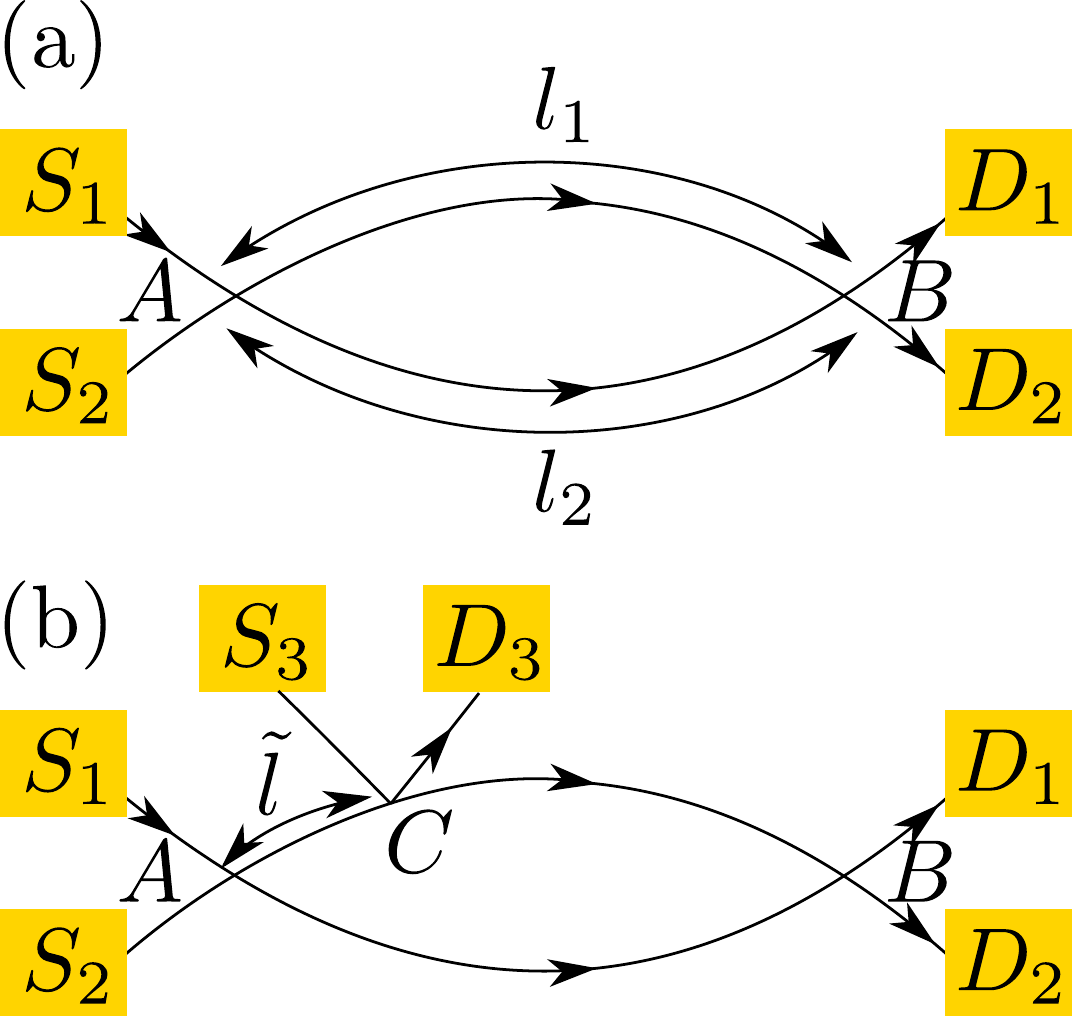}
\end{center}
\caption{
Illustration of the Mach-Zehnder interferometers (MZIs) under
study. Chiral channels are represented by full lines leading from the
sources ($S_1$ --- biased and $S_2$ --- grounded) to the drains ($D_1$ and
$D_2$). Inter-edge tunneling takes place at intersection points. (a) A
standard MZI with arms 1 and 2 of lengths $l_1$ and $l_2$,
respectively. (b) The dangling end at $C$ (leading to $D_3$) serves as
an absorber replacing the ``lousy'' bomb. 
 } \label{setup}
\end{figure}

Here we analyze the correlations of transport through an
e-MZI with a leaking edge. This is an electronic manifestation of a
variant of the EV-bomb where the leaky-edge corresponds to an absorber
instead of a bomb~\cite{Mitchison:2001}. In the particle-like limit of
this device, the probability  of a particle being absorbed and
transmitted to the drains at the same time is zero. Such  correlations
in the case of many-particles will yield a non-vanishing result. This
signifies the fact that the bomb may ``explode'' even if a signal is
detected at the interferometer's dark port. Employing  a wave-like
scattering matrix formulation, we compute  the experimentally
measurable many-body correlator  and compare to  two limiting cases
(single particle impinging vs. a large influx of particles).  Subsequently, we find the conditions for manifesting  the wave-particle duality, and specifically obtaining the EV physics, in the context of many-body electronic system. 
                                                                     
\section{System}
We consider a standard e-MZI geometry where particles are injected from the source $S_1$ and eventually detected at the
drains, $D_1$ and $D_2$ [see Fig.~\ref{setup}(a)]. Note that all channels are
chiral, i.e. particles may move only in the direction of the
arrow.
The evolution of an injected wave packet through the setup is described by considering incoming scattering states from the various sources that are labeled by their quantum number $k$. 
Schematically, the state of a particle injected from $S_1$, after
passing through beam-splitter $A$ at position $x=0$, is described by
$\ket{i}=r^{\phantom{\dagger}}_A\ket{1}+t^{\phantom{\dagger}}_A\ket{2}$,
where $r^{\phantom{\dagger}}_A$ and $t^{\phantom{\dagger}}_A$ are the
reflection and transmission amplitudes~\cite{Note1} corresponding to
beam-splitter $A$, and $\ket{1},\ket{2}$ are the scattering states
corresponding to the upper and lower e-MZI arms. 
Similarly, the beam splitter $B$ is characterized by  reflection
and transmission amplitudes $r_B$ and $t_B$, respectively. 
Between the beam splitters $A$ and $B$, orbital phases are accumulated along arm 1 and arm 2, i.e. $e^{ik 
  l_1}$ and $e^{ik l_2}$, respectively. Additionally, for charged particles in the presence of a magnetic field, the relative phase of the two respective trajectories includes  an Aharonov-Bohm phase $\Phi_{\rm AB}\equiv 2 \pi \frac{\Phi}{\Phi_0}$, where $\Phi_0$ is a quantum of flux.
With a proper gauge choice, we reabsorb these phases in an extra phase
shift of the transmission coefficient of $t_B \to t_B e^{i\phi}$, with
the interference phase  $\phi=\phi^{\phantom{\dagger}}_B\equiv
k(l_2-l_1) +\Phi_{\rm AB}$.

We incorporate a semi-transparent absrober on the arm-1 of the e-MZI using an additional beam-splitter $C$ at position $0<\tilde{l} < l_1$ [see Fig.\ref{setup}(b)]. The propagation of an impinging particle is thus
modified: the particle may exit the MZI through arm $3$ and reach drain $D_3$. The effect of this extra beam splitter evolves the scattering state component in arm-1,  $\ket{1} \to  r^{\phantom{\dagger}}_C\ket{1} + t^{\phantom{\dagger}}_C\ket{3}$. This process is commonly referred to as \textit{partial-collapse} and has been studied in the context of qubit-uncollapse~\cite{Korotkov:2007,Katz08} and null weak values~\cite{Zilberberg:2013,zilberbergCrypta,zilberberg2014standard}.

This schematic evolution through the e-MZI can be conveniently recast in a scattering matrix formulation, i.e., we can write the state of a particle in the interferometer in second quantization, with an annihilation operator
\begin{gather}\label{solution}
\psi_{km}(x)=e^{ikx} \left\{
\begin{array}{ll}
    a^{\phantom{\dagger}}_{km}, & \hbox{$x<0;$} \\
    b^{\phantom{\dagger}}_{km}, & \hbox{$0<x<\tilde{l};$} \\
    c^{\phantom{\dagger}}_{km}, & \hbox{$\tilde{l}<x<l_2;$} \\
    d^{\phantom{\dagger}}_{km}, & \hbox{$l_m<x.$} \\
\end{array}
\right.
\end{gather}
Here $m=1,2,3$ labels the different device arms and we assume arbitrarily that $l_2<l_1$. 
The operators $a_{km}$, $b_{km}$, $c_{km}$, $d_{km}$ are the
annihilation operators of momentum eigenstates in the different
sectors of the interferometer. They can be arranged in vectors
$\mathbf{a}_k$, $\mathbf{b}_k$, $\mathbf{c}_k$, $\mathbf{d}_k$,
labeled by the arm-index $m$, and are related by scattering
matrices  describing the effects of beam splitters via
\begin{align}
\mathbf{b}^{\phantom{\dagger}}_k=\mathcal{S}^{\phantom{\dagger}}_A
\mathbf{a}^{\phantom{\dagger}}_k,\, \mathbf{c}^{\phantom{\dagger}}_k=\mathcal{S}^{\phantom{\dagger}}_C
\mathbf{b}^{\phantom{\dagger}}_k,\, \mathbf{d}^{\phantom{\dagger}}_k=\mathcal{S}^{\phantom{\dagger}}_B
\mathbf{c}^{\phantom{\dagger}}_k, \label{scatterer}
\end{align}
 with 
\begin{align}
\mathcal{S}^{\phantom{\dagger}}_i &= \left(%
\begin{array}{ccc}
  r^{\phantom{\dagger}}_i & t^{\phantom{\dagger}}_i & 0 \\
  -t_i^* & r^{\phantom{\dagger}}_i & 0 \\
  0 & 0 & 1 
\end{array}%
\right) \,\, ; \,\, i=A,B, \\
\mathcal{S}^{\phantom{\dagger}}_C &= \left(%
\begin{array}{ccc}
  r^{\phantom{\dagger}}_C & 0 & -t^{\phantom{\dagger}}_C \\
  0 & 1 & 0 \\
  t^{\phantom{\dagger}}_C & 0 & r_C^* 
\end{array} \right). \nonumber
\end{align}


\section{Single-particle limit}
As a first step we analyze the the effect of the extra beam splitter
$C$ using a schematic single-particle formulation. We assume that
the incoming state is labeled by the quantum number $k$, which, for
clarity we omit in the notation below. 
In the absence of the leakage port, the probability to measure the particle in drain $D_1$ is
$P_0(i\rightarrow D_1)=|\bra{D_1}\ket{i}|^2$, where $\ket{i}=r^{\phantom{\dagger}}_A\ket{1}+t^{\phantom{\dagger}}_A\ket{2} $ includes
the effect of beam splitter $A$, and  we have defined $\ket{D_1}=r^{\phantom{\dagger}}_B\ket{1}+t^{\phantom{\dagger}}_B
e^{i\phi}\ket{2}$ to include the effect of
beam splitter $B$  and the subsequent detection in $D_1$.  We have used the
subscript ${\cdot}_0$ to denote the probability in the absence of a
leakage port. We obtain for the setup of  Fig.~\ref{setup}(a),
$
	P_0(i\rightarrow
        D_1)=|r^{\phantom{\dagger}}_A|^2|r^{\phantom{\dagger}}_B|^2+|t^{\phantom{\dagger}}_A|^2|t^{\phantom{\dagger}}_B|^2
        +2|r^{\phantom{\dagger}}_A r^{\phantom{\dagger}}_B t^{\phantom{\dagger}}_A t^{\phantom{\dagger}}_B|\cos(\phi+\phi^{\phantom{\dagger}}_T)$,
where $\phi^{\phantom{\dagger}}_T=\arg(r^{\phantom{\dagger}}_A
r^{\star}_B t^{\phantom{\dagger}}_A t^{\star}_B)$. We think of the
state of the propagating electron as a superposition of quibit states,
$\ket{1}$, $\ket{2}$.

Introducing the beam-splitter $C$ on arm 1, allows the state $\ket{1}$ to
``leak out'' (partial-collapse) to through branch 3 with probability
$|t^{\phantom{\dagger}}_C|^2$ [cf. Fig.~\ref{setup}(b)]. 
The probability  to reach drain $D_3$ is therefore,
\begin{align}
P(i\rightarrow D_3)=|r^{\phantom{\dagger}}_A|^2|t^{\phantom{\dagger}}_C|^2\,. 
\label{probD3}
\end{align}
Upon detection of the injected electron in $D_3$, we declare the interference experiment void. In such a "partial collapse"  the state $\ket{1}$ is 
projected out of the space spanned by $\ket{1}$ and $\ket{2}$. If such
a projection-out does not take place (i.e. the electron is \emph{not}
detected in $D_3$), the original qubit state is
rotated by the measurement's back-action into
$\ket{i_C}=(1/\tilde{\mathcal{N}})\left(r^{\phantom{\dagger}}_A
  r^{\phantom{\dagger}}_C\ket{1}+t^{\phantom{\dagger}}_A\ket{2}\right)$
with normalization $\tilde{\mathcal{N}}=\sqrt{1-P(i\rightarrow D_3)}$.
Consequently, the probability for the particle to subsequently arrive
in drain $D_1$ is $P(i_C \rightarrow D_1)P(\overline{i\rightarrow
  D_3})$, where by overline we denote the complementary event,
i.e. $P(\overline{i\rightarrow D_3})=1-P(i\rightarrow D_3)$.  
Note that $P(i_C \rightarrow D_1)$ can be written using the conditional probability $P(i \rightarrow D_1 \,|\, \overline{i\rightarrow D_3})$.
As a result we obtain that the particle would reach drain $D_1$ with the joint probability 
\begin{align}
\label{collapseProb}
&P(i\rightarrow D_1)	= P(i\rightarrow D_1 , \overline{i\rightarrow D_3})
=|r^{\phantom{\dagger}}_A|^2|r^{\phantom{\dagger}}_B|^2|r^{\phantom{\dagger}}_C|^2
\\ &+|t^{\phantom{\dagger}}_A|^2|t^{\phantom{\dagger}}_B|^2
+2|r^{\phantom{\dagger}}_C||r^{\phantom{\dagger}}_A r^{\phantom{\dagger}}_B t^{\phantom{\dagger}}_A t^{\phantom{\dagger}}_B|\cos(\phi+\phi^{\phantom{\dagger}}_T+\phi^{\phantom{\dagger}}_C)\,,\nonumber	
\end{align}
where $\phi^{\phantom{\dagger}}_C=\arg(r^{\phantom{\dagger}}_C)$. Note that due to causality $P(i\rightarrow D_1)=P(i\rightarrow D_1 , \overline{i\rightarrow D_3})$ and similarly 
\begin{align}
P(i\rightarrow D_1 , i\rightarrow D_3)=0\,. 
\label{holyGrail}
\end{align}
The fact that $P(i \rightarrow D_1 )\neq P_0(i \rightarrow D_1 )$ can
be used to detect the presence of the leakage port.  
Specifically, if the MZI is tuned to have $P_0(i \rightarrow D_1
)=0$, the detection of a
particle at $D_1$ in any single realization of the experiment indicates the presence of the leakage port without the
particle having leaked out. If the particle is not detected at $D_1$, no
conclusion on the presence of a leakage channel can be drawn. This is a manifestation of the EV-bomb detection scheme. 

It is instructive to recover the results of this single particle analysis in
the scattering matrix formalism, which provides the basis to analyze
the statistical many-body effects in the following section.
In the scattering matrix formalism we consider the injection of a {\it single particle} (in
the scattering state $k$) into the system,
i.e. $\ket{i_k}=a^\dagger_{k,1}\ket{0}$. The detection of the particle
in $D_{1(3)}$ is described by the projection operator $\Pi_{D_1(D_3)} \equiv
d^{\dag}_{k,1(3)}d^{\phantom{\dagger}}_{k,1(3)}$. From Eq.~(\ref{scatterer}), the probabilities for the
injected particle to reach $D_1$ or $D_3$ are
\begin{align}
	\label{currentD1}
	P(i_k \rightarrow D_1) =\bra{0}a^{\phantom{\dagger}}_{k,1}
        d^{\dag}_{k,1}d^{\phantom{\dagger}}_{k,1}
        a^\dagger_{k,1}\ket{0}&
=\mathcal{A}_{11},\\
	P(i_k \rightarrow D_3)= \bra{0}a^{\phantom{\dagger}}_{k,1}
        d^{\dag}_{k,3}d^{\phantom{\dagger}}_{k,3}
        a^\dagger_{k,1}\ket{0} &
        =\mathcal{B}_{11}\,,
	\label{currentD3}
\end{align}
 where we have introduced the  quantities $\mathcal{A}_{ij} \equiv
(\mathcal{S}_A^{\dag}\mathcal{S}_B^{\dag}\mathcal{S}_C^{\dag})_{i1}(\mathcal{S}_C\mathcal{S}_B\mathcal{S}_A)_{1j}$,
$ \mathcal{B}_{ij}\equiv (\mathcal{S}_A^{\dag}\mathcal{S}_B^{\dag}\mathcal{S}_C^{\dag})_{i3}(\mathcal{S}_C\mathcal{S}_B\mathcal{S}_A)_{3j}$.
 Indeed, an explicit evaluation of $\mathcal{A}_{11}$ and
 $\mathcal{B}_{11}$ yields, for Eqs.~\eqref{currentD1} and
 \eqref{currentD3} exactly the same expressions as
 Eqs.~\eqref{collapseProb} and \eqref{probD3}, respectively.  

Additionally, the joint probability of detecting a particle at $D_1$ and $D_3$
is given by 
\begin{multline}
P(i\rightarrow D_1 , i\rightarrow D_3) 
\bra{0}a^{\phantom{\dagger}}_{k,1}
   d^{\dag}_{k,3}d^{\phantom{\dagger}}_{k,3}d^{\dag}_{k,1}d^{\phantom{\dagger}}_{k,1}
   a^\dagger_{k,1}\ket{0}\\
= \sum_{\beta=1}^3
    \mathcal{A}_{1\beta}\mathcal{B}_{\beta 1}\equiv 0\,,  
	\label{crosscurrent}
\end{multline}
where, when the incoming state is of a single particle, we recover
the result in Eq.~\eqref{holyGrail}.

The results of this section describe experiments where a single
particle is injected into the interferometer. While this is possible
in quantum optics, it does not represent the typical experimental
conditions of electronic devices. Single-particle sources have been only recently reported in some specifically designed experimental architectures~\cite{Dubois:2013}. Since many-electron physics is an essential part of quantum reality, we next analyze this limit.

\section{Many-body conditional correlations}

In a typical experiment with e-MZI, particles are injected into the
source from a
voltage biased reservoir, and  are collected in the drain
over a macroscopically long time. 
This being the case, only
statistical quantities averaged over a many-particle ensemble are
accessible, and the signals at the detector correspond to statistical
averages of the source-drain transition probabilities computed in the
previous section.
Specifically, for an e-MZI with a voltage bias $eV$ at $S_1$, the
measured current at $D_1$ is given by the rate of electrons reaching this drain
out of the total rate, $eV/\hbar$, of electrons impinging from the source. The currents through the device are therefore statistical
probabilities for an impinging electron to reach the various drains, and are
precisely given in terms of the probabilities calculated in the single-particle picture above: the current at drain $j$ will be given by $I_j=
(e^2/h) P(i\rightarrow D_j) V$.
When the signal in $D_1$ is collected over a large
number of particles, any outcome of the IFM-experiment
would have a macroscopic leakage of particles in $D_3$ even if the e-MZI is tuned to have a vanishing
current in the absence of the port $D_3$. Hence, in the original formulation
of the problem with the bomb, the bomb would necessarily explode. In short, under the above conditions the detection of the current at $D_3$ does not constitute an  uncontested manifestation of IFM. 

Can, and under what conditions, an electronic MZI setup reproduce the original EV
bomb measurement scheme? In order to clarify this we focus on the difference between the single-particle results and the
many-particle statistical averages relevant for experiments, which
appears when dealing with joint probabilities. 

This is clearly
demonstrated considering, e.g., the statistical joint probability of detecting
particles at drain $D_1$ and $D_3$. 
In order to relate such a  joint probability with a quantity directly accessible in experiments,
we next study the current-current correlations in
a {\it many-body} (albeit non-interacting) system. 
We assume that a voltage bias $V$ is applied to the source $S_1$,
which is held at temperature $T$. 
For a system with linear dispersion relation, the
current operator is
$\hat{I}_i(x,t)=ev:\psi_i^{\dag}(x,t)\psi_i(x,t):$, where
$\psi_i(x,t)$ is the annihilation operator in the $i$-th arm, and
the normal order operator, $:\,:$, indicates the subtraction of the
mean equilibrium contribution.

We consider the cross-current correlation defined by 
\begin{equation}
F_{1,3} \equiv \frac{h^2}{e^4 V^4 \tau} \int_{-\tau/2}^{\tau/2}dt 
\ave{\hat{I}_3(x_0,t) \hat{I}_1(x,0)},
\end{equation}
 where $\tau$ is an infrared cut-off, $\tau\gg\frac{L}{v}$, and
$x_0,x>l_1$. 
Importantly, since the average  current is related to the electron
transfer probability by the factor $e^2V/h$, the prefactor in the
definition of $F_{1,3}$ allows us to directly compare this correlator with
the averaged  joint probability of detecting electrons at drain $D_1$
and $D_3$ [cf.~Eq.~\eqref{crosscurrent}].

Using Wick's theorem, the fact that all ohmic contacts are grounded apart from $S_1$ which is at $eV$, the identity $f_\alpha(1-f_\beta)=\frac{1}{2}\left[f_\alpha(1-f_\alpha)+f_\beta(1-f_\beta)+(f_\alpha-f_\beta)+(f_\alpha-f_\beta)^2\right]$ where $f$ is the Fermi-Dirac distribution, and the limit of $\tau \gg L/v$, we obtain

\begin{widetext}
\begin{align}
&F_{1,3} \equiv F_\infty (\alpha,\Delta \tilde{L}) - F_N (\alpha,\Delta \tilde{L}) = \frac{1}{ \alpha}|r_A|^2|t_C|^2\Bigg[\alpha \left(|t_A|^2 |t_B|^2+|r_A|^2 |r_B|^2 |r_C|^2\right)+2K(\alpha,\Delta \tilde{L})|t_A t_B r_A r_B r_C|\cos[\Phi(\alpha,\Delta \tilde{L},\Phi_{\rm AB})] \Bigg]\nonumber\\
&-\frac{1}{\alpha N} |r_A|^2|t_C|^2\Bigg[L(\alpha)\left(|t_A|^2 |t_B|^2+|r_A|^2 |r_B|^2 |r_C|^2\right)+2M(\alpha,\Delta \tilde{L})|t_A t_B r_A r_B r_C|\cos[\Phi(\alpha,\Delta \tilde{L},\Phi_{\rm AB})] \Bigg]\, .
\label{condEq}
\end{align}
\end{widetext}
where $F_\infty$ and $F_N$ are functions of the  dimensionless
parameters $\alpha=eV \beta/(2\pi)$, $\Delta \tilde{L}=\pi (l_2-l_1)/(\hbar \beta v)$, and $N=eV\tau/(2\pi\hbar)$. Here $\beta=1/(k_B T)$ is the inverse temperature. We have also introduced the functions 
$K(\alpha,\Delta \tilde{L})=\sin[\alpha\Delta \tilde{L}]/\sinh[\Delta \tilde{L}]$, $L(\alpha)=(\pi \alpha \coth[\pi
\alpha]-1)/\pi $, $M(\alpha,\Delta \tilde{L})=(\pi \sin[\alpha \Delta
L]\coth[\pi \alpha]-\Delta \tilde{L} \cos[\alpha\Delta \tilde{L}])/ (\pi \sinh[\Delta
L])$, and $\Phi(\alpha,\Delta \tilde{L},\Phi_{\rm AB})=\Phi_{\rm AB}+\phi^{\phantom{\dagger}}_T+\phi^{\phantom{\dagger}}_C+\alpha \Delta \tilde{L}$.

\begin{figure}
\begin{center}
\includegraphics[width=\columnwidth]{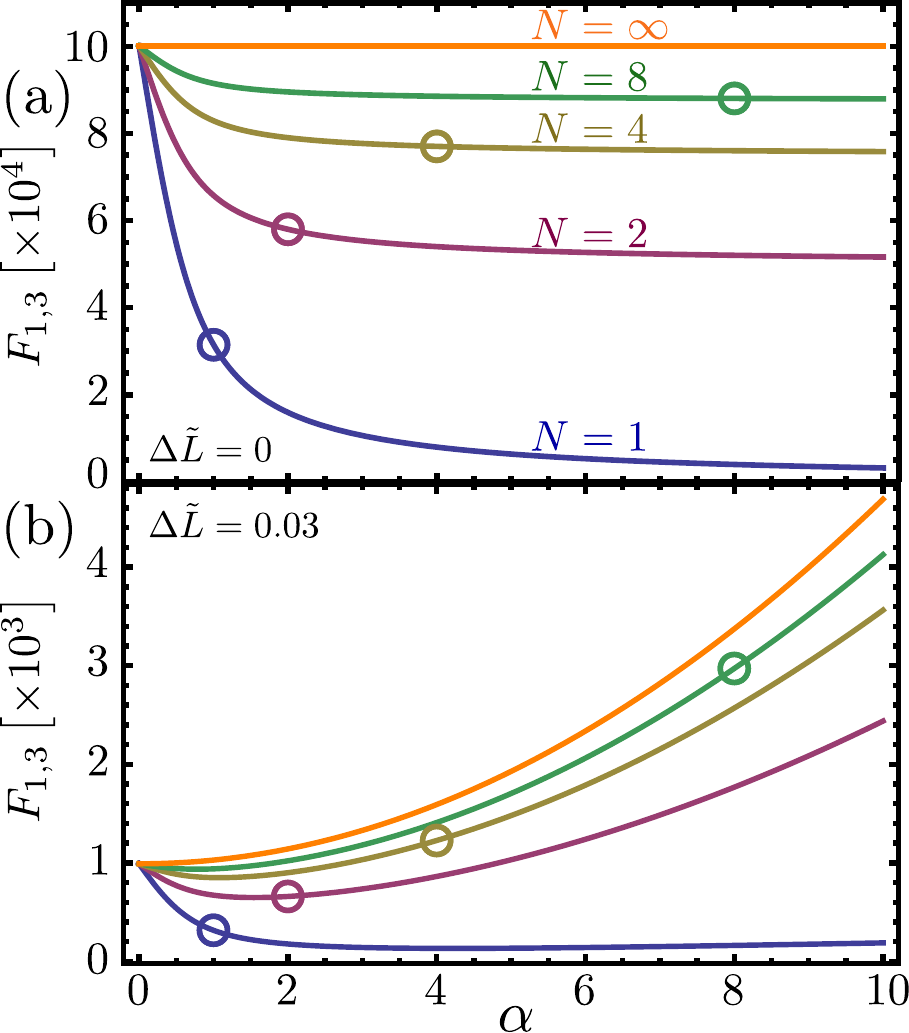}
\end{center}
\caption{ The many-body cross-current correlator as a function of $\alpha$ [cf.~\eqref{condEq}]. Here, we have taken $|t_C|^2=0.3$,$|r_A|^2=|r_B|^2=0.5$ and 
$\Phi_{\rm AB}+\phi^{\dagger}_T+\phi^{\dagger}_C=\pi$. For a temperature of $T=10$mK the
  parameter $\alpha$ corresponds to realistic bias values of up to
  $\sim 54\mu$V. The different plots correspond to different values of
  $N=1\ldots 10$. As a function of $\alpha$ for a fixed $T$, $\tau$ should be changed in order to keep $N$ constant, i.e. 
	$\tau= N\frac{2\pi\hbar}{eV}=N\frac{\hbar\beta}{\alpha}\sim 7.63823\times 10^{-10}(N/\alpha)$[second]. We mark by circles the point $\alpha\equiv N$ as the threshold for which our assumption $\tau \gg L/v$ breaks for existing electronic interferometers~\cite{Neder:2007}. 
	(a) The case of $\Delta \tilde{L} \rightarrow 0$. 
	(b) The case of $\Delta \tilde{L} =0.03$ where dephasing affects both the classical and quantum correlators (due to varying interference lengths per wavenumber). Nonetheless, the single-particle limit remains unaffected as expected from Eq.~\eqref{holyGrail}. }\label{fig2}
\end{figure}

Before discussing the implication of this result, it is instructive to
contrast 
the many-body conditional correlator to purely classical correlations of
an ensemble of statistically independent impinging electrons.
In the latter case, we obtain the statistical average of a joint signal at port $D_1$ and $D_3$:
\begin{multline}
\label{statisticalMeans}
	 \tilde{P}(i\rightarrow D_1 , i\rightarrow D_3)=P(i\rightarrow D_1)P(i\rightarrow D_3)\\
=|r^{\phantom{\dagger}}_A|^2|t^{\phantom{\dagger}}_C|^2\Big[|r^{\phantom{\dagger}}_A|^2|r^{\phantom{\dagger}}_B|^2|r^{\phantom{\dagger}}_C|^2+|t^{\phantom{\dagger}}_A|^2|t^{\phantom{\dagger}}_B|^2\\
 +2|r^{\phantom{\dagger}}_C||r^{\phantom{\dagger}}_A r^{\phantom{\dagger}}_B t^{\phantom{\dagger}}_A t^{\phantom{\dagger}}_B|\cos(\phi+\phi^{\phantom{\dagger}}_T+\phi^{\phantom{\dagger}}_C)\Big]\,.	
\end{multline}
For a better comparisson with the  full many-body results that include the effect of
averaging over a statistical ensamble due to termal fluctuations, as
well as out-of equilibrium voltage bias, one can further average
over a density matrix, $\rho$, that describes and ensemble of initial states. 
For example, assuming that a voltage bias $V$ is applied to the source $S_1$,
which is held at temperature $T$, the state of the impinging electrons
is described by $\rho=(1/L)
\sum_k [f(\hbar vk-eV)-f(\hbar vk)]a_{k1}^{\dag}a_{k1}$, where $f(x)$
is the 
Fermi-Dirac distribution, and the system length, $L$, is taken to be
la largest length scale in the problem. 
When averaged over the initial density matrix, the ``classical''
correlations in Eq.~(\ref{statisticalMeans}) yield
\begin{multline}
\label{MBstatMean}
\tilde{P}(i\rightarrow D_1 , i\rightarrow D_3)=|r^{\phantom{\dagger}}_A|^2|t^{\phantom{\dagger}}_C|^2\left[|r^{\phantom{\dagger}}_A|^2|r^{\phantom{\dagger}}_B|^2|r^{\phantom{\dagger}}_C|^2+|t^{\phantom{\dagger}}_A|^2|t^{\phantom{\dagger}}_B|^2\right]\\
+2\frac{K(\alpha) }{\alpha}|r^{\phantom{\dagger}}_C||r^{\phantom{\dagger}}_A r^{\phantom{\dagger}}_B t^{\phantom{\dagger}}_A t^{\phantom{\dagger}}_B|\cos\left[\Phi(\alpha,\Delta \tilde{L},\Phi_{\rm AB})\right]\,.
\end{multline}

Comparing the statistical probability analysis in
Eq.~\eqref{MBstatMean} with the many-body joint correlation in
Eq.~(\ref{condEq}), we obtain that $\tilde{P} (i\rightarrow D_1 ,
i\rightarrow D_3)=F_\infty$, which is the dominant contribution of
$F_{1,3}$ in the zero-frequency DC-limit.
Indeed, this represents the well-known fact that $\lim_{\tau \to \infty} \int_{-\tau/2}^{\tau_2} dt  \langle \hat{I}_3(x_0,t) \hat{I}_1(x,0) \rangle
=\langle \hat{I}_3(x_0,t) \rangle \langle \hat{I}_1(x,0) \rangle \tau$.
Similarly, a standard analysis of current-current
correlations~\cite{Blanter:2000} singles out the non-trivial correlations in the
cross-current noise $S_{1,3} \equiv \lim_{\tau \to \infty} \int_{-\tau/2}^{\tau_2} dt  ( \langle \hat{I}_3(x_0,t) \hat{I}_1(x,0) \rangle -
  \langle \hat{I}_3(x_0,t) \rangle \langle \hat{I}_1(x,0) \rangle )$.
These non-trivial contributions are encoded in the term $F_N =
S_{1,3}/ ( \hat{I}_0^2 \tau )$ of the
many-body cross-current correlation in Eq.~(\ref{condEq}). Technically it
corresponds to a particle-hole loop
contribution. 

While at low frequencies, $F_\infty$ is the dominant contribution to 
cross-current correlations, Eq.~(\ref{condEq}) clearly shows how, for
measurements averaged over a finite time, the effects of
$F_\infty$ and $F_N$ are competing. In fact,
they become of the same order for short measurements times, such that
the average currents are comparable with their fluctuations,
i.e., $\langle \hat{I}_3(x_0,t) \rangle \langle \hat{I}_1(x,0) \rangle \tau\sim  S_{1,3}$.
In particular, one expects that in the limit where the average number of particles
in the interferometer is $\sim 1$ during the measurement time $\tau$,
these two terms 
cancel each other, and we can recover the single-particle result
of Eq.~(\ref{holyGrail}). 
By estimating the average number of electrons impinging on the e-MZI during the
measurement time by $N=\frac{eV\tau}{2\pi\hbar}$, we are in the
position of interpreting the cross-current correlator in terms of
a crossover between single-particle quantum-mechanical correlations
and classical statistical correlations.

Fig.~\ref{fig2}(a) depicts the cross-current correlations 
as  function of the voltage bias, $\alpha$, measured in units of temperature, for different values of $N$. 
For any value of $N$, at
$\alpha \lesssim 4$, thermal fluctuations dominate over
the quantum ones, and the correlations will ultimately reduce to those
of classical
waves.
For large $\alpha$, upon decreasing $N$, $F_{1,3}$ decreases, and for
$N \sim 1$ it is essentially vanishing, i.e., we obtain $F_{1,3} \ll F_\infty$ which signals
that quantum correlations are important. 
Note that Eq.~\eqref{condEq}, depicted in Fig.~\ref{fig2}, is valid for
$\tau \gg L/v$. Recall that as a function of $\alpha$ for a fixed temperature $T$,
$\tau$ changes in order to keep a constant $N$, i.e. $\tau=
  N\frac{2\pi\hbar}{eV}=N\frac{\hbar\beta}{\alpha}\sim 7.63823\times
  10^{-10}(N/\alpha)$[second], where we considered $T=10$mK. 
Taking experimental values of existing electronic interferometers, $L\sim 10\mu$m and $v\sim 2-6\times 10^{-4}$m/s~\cite{Neder:2007}, we mark the point $\alpha\equiv N$ as a threshold beyond which our prediction no longer holds. As such, in order to reach the limit of single-particle demonstration of IFM, one should construct smaller interferometers or generate higher edge mobility. Alternatively, one could consider single-particle injection on top of a Fermi sea~\cite{Dubois:2013}, but this is beyond the scope of our analysis.
 
In Fig.~\ref{fig2}(b), we see the effect of a finite $\Delta
\tilde{L}$. As each wavenumber experiences a slightly different
interference path, both the classical and quantum many-body
correlations are affected by averaging over many wavenumbers. As a
result, when many particles are considered [Eq.~\eqref{condEq}] the
result moves further away from the single-particle limit of
Eq.~\eqref{holyGrail} reflecting this effective
dephasing. Nonetheless, in the limit of short pulses, $N=1$, the
correlator yields an outcome that agrees with the single-particle picture.

\section{Conclusions}

The main focus of this study is the assessment of feasible detection
of IFM in a genuine many-body electronic system. To this goal, we have analyzed an electronic MZI with a leakage port located on one of the
interferometer arms, which servs as an experimentally viable implementation of the
EV-bomb gedanken experiment. We considered the typical experimental
settings when an ensemble of particles is injected in the
interferometer, i.e., the current in the interferometer yields a
statistically averaged signal. 
We analyzed the cross-current correlation at the dark and leakage
ports, which is vanishing in the single-particle original proposal of
the experiment, but remains generally finite in the many-particle statistical
implementation. This has allowed us to identify the parameters' regime (voltage bias, temperature)
for which the many-body correlations approach the single-particle
result. 
We find the regime where the wave-particle duality emerges is
lies just at the frontiers of actual experiments with electronic MZIs, where the
main limitations are due to the size of the interferometer and the
mobility of the electrons at the edges of a Hall bar.

In summary, our results show that the detection of IFM in a many-body
electronic system seems to involve two competing facets that need to
be dealt with: IFM a-la Elitsur-Vaidman requires to deal with particles (that, in principle, can be pin-pointed to a specific spatial coordinate); at the same time, the setup employed is an interferometer, which invokes the wave character of the quantum object. One thus needs to fine-tune the system to zoom on a regime where particle-wave duality is manifest.  Our analysis might trigger experiments with single-electron biased MZIs, where this physics may be elucidated.

\acknowledgments

This work has been supported by the Swiss National Science Foundation, the German-Israel Foundation (GIF),
Deutsche Forschungsgemeinschaft (DFG) grants and RO 2247/8-1 and RO 4710/1-1, and the Israel Science Foundation (ISF).


\begin{thebibliography}{30}
\expandafter\ifx\csname natexlab\endcsname\relax\def\natexlab#1{#1}\fi
\expandafter\ifx\csname bibnamefont\endcsname\relax
  \def\bibnamefont#1{#1}\fi
\expandafter\ifx\csname bibfnamefont\endcsname\relax
  \def\bibfnamefont#1{#1}\fi
\expandafter\ifx\csname citenamefont\endcsname\relax
  \def\citenamefont#1{#1}\fi
\expandafter\ifx\csname url\endcsname\relax
  \def\url#1{\texttt{#1}}\fi
\expandafter\ifx\csname urlprefix\endcsname\relax\def\urlprefix{URL }\fi
\providecommand{\bibinfo}[2]{#2}
\providecommand{\eprint}[2][]{\url{#2}}

\bibitem[{\citenamefont{Haroche}(2013)}]{haroche_nobel_2013}
\bibinfo{author}{\bibfnamefont{S.}~\bibnamefont{Haroche}},
  \bibinfo{journal}{Rev. Mod. Phys.} \textbf{\bibinfo{volume}{85}},
  \bibinfo{pages}{1083} (\bibinfo{year}{2013}).

\bibitem[{\citenamefont{Elitzur and Vaidman}(1993)}]{Elitzur:1993}
\bibinfo{author}{\bibfnamefont{A.}~\bibnamefont{Elitzur}} \bibnamefont{and}
  \bibinfo{author}{\bibfnamefont{L.}~\bibnamefont{Vaidman}},
  \bibinfo{journal}{Quant. Opt.} \textbf{\bibinfo{volume}{6}},
  \bibinfo{pages}{119} (\bibinfo{year}{1993}).

\bibitem[{\citenamefont{Kwiat et~al.}(1995)\citenamefont{Kwiat, Weinfurter,
  Herzog, Zeilinger, and Kasevich}}]{Kwiat:1995}
\bibinfo{author}{\bibfnamefont{P.}~\bibnamefont{Kwiat}},
  \bibinfo{author}{\bibfnamefont{H.}~\bibnamefont{Weinfurter}},
  \bibinfo{author}{\bibfnamefont{T.}~\bibnamefont{Herzog}},
  \bibinfo{author}{\bibfnamefont{A.}~\bibnamefont{Zeilinger}},
  \bibnamefont{and} \bibinfo{author}{\bibfnamefont{M.~A.}
  \bibnamefont{Kasevich}}, \bibinfo{journal}{Phys. Rev. Lett.}
  \textbf{\bibinfo{volume}{74}}, \bibinfo{pages}{4763} (\bibinfo{year}{1995}).

\bibitem[{\citenamefont{du~Marchie~van Voorthuysen}(1996)}]{Voorthuysen:1996}
\bibinfo{author}{\bibfnamefont{E.~H.} \bibnamefont{du~Marchie~van
  Voorthuysen}}, \bibinfo{journal}{American Journal of Physics}
  \textbf{\bibinfo{volume}{64}} (\bibinfo{year}{1996}).

\bibitem[{\citenamefont{Hafner and Summhammer}(1997)}]{Hafner:1997}
\bibinfo{author}{\bibfnamefont{M.}~\bibnamefont{Hafner}} \bibnamefont{and}
  \bibinfo{author}{\bibfnamefont{J.}~\bibnamefont{Summhammer}},
  \bibinfo{journal}{Physics Letters A} \textbf{\bibinfo{volume}{235}},
  \bibinfo{pages}{563 } (\bibinfo{year}{1997}), ISSN \bibinfo{issn}{0375-9601}.

\bibitem[{\citenamefont{Tsegaye et~al.}(1998)\citenamefont{Tsegaye, Goobar,
  Karlsson, Bj\"ork, Loh, and Lim}}]{Tsegaye:1998}
\bibinfo{author}{\bibfnamefont{T.}~\bibnamefont{Tsegaye}},
  \bibinfo{author}{\bibfnamefont{E.}~\bibnamefont{Goobar}},
  \bibinfo{author}{\bibfnamefont{A.}~\bibnamefont{Karlsson}},
  \bibinfo{author}{\bibfnamefont{G.}~\bibnamefont{Bj\"ork}},
  \bibinfo{author}{\bibfnamefont{M.~Y.} \bibnamefont{Loh}}, \bibnamefont{and}
  \bibinfo{author}{\bibfnamefont{K.~H.} \bibnamefont{Lim}},
  \bibinfo{journal}{Phys. Rev. A} \textbf{\bibinfo{volume}{57}},
  \bibinfo{pages}{3987} (\bibinfo{year}{1998}).

\bibitem[{\citenamefont{Kwiat et~al.}(1999)\citenamefont{Kwiat, White,
  Mitchell, Nairz, Weihs, Weinfurter, and Zeilinger}}]{Kwiat:1999}
\bibinfo{author}{\bibfnamefont{P.~G.} \bibnamefont{Kwiat}},
  \bibinfo{author}{\bibfnamefont{A.~G.} \bibnamefont{White}},
  \bibinfo{author}{\bibfnamefont{J.~R.} \bibnamefont{Mitchell}},
  \bibinfo{author}{\bibfnamefont{O.}~\bibnamefont{Nairz}},
  \bibinfo{author}{\bibfnamefont{G.}~\bibnamefont{Weihs}},
  \bibinfo{author}{\bibfnamefont{H.}~\bibnamefont{Weinfurter}},
  \bibnamefont{and}
  \bibinfo{author}{\bibfnamefont{A.}~\bibnamefont{Zeilinger}},
  \bibinfo{journal}{Phys. Rev. Lett.} \textbf{\bibinfo{volume}{83}},
  \bibinfo{pages}{4725} (\bibinfo{year}{1999}).

\bibitem[{\citenamefont{Jang}(1999)}]{Jang:1999}
\bibinfo{author}{\bibfnamefont{J.-S.} \bibnamefont{Jang}},
  \bibinfo{journal}{Phys. Rev. A} \textbf{\bibinfo{volume}{59}},
  \bibinfo{pages}{2322} (\bibinfo{year}{1999}).

\bibitem[{\citenamefont{Hosten et~al.}(2006)\citenamefont{Hosten, Rakher,
  Barreiro, Peters, Peters, and Kwiat}}]{Hosten:2006}
\bibinfo{author}{\bibfnamefont{O.}~\bibnamefont{Hosten}},
  \bibinfo{author}{\bibfnamefont{M.~T.} \bibnamefont{Rakher}},
  \bibinfo{author}{\bibfnamefont{J.~T.} \bibnamefont{Barreiro}},
  \bibinfo{author}{\bibfnamefont{N.~A.} \bibnamefont{Peters}},
  \bibinfo{author}{\bibfnamefont{N.~A.} \bibnamefont{Peters}},
  \bibnamefont{and} \bibinfo{author}{\bibfnamefont{P.~G.} \bibnamefont{Kwiat}},
  \bibinfo{journal}{Nature} \textbf{\bibinfo{volume}{439}},
  \bibinfo{pages}{949} (\bibinfo{year}{2006}).

\bibitem[{\citenamefont{Wolfgramm et~al.}(2011)\citenamefont{Wolfgramm,
  de~Icaza~Astiz, Beduini, Cer\`e, and Mitchell}}]{Wolfgramm:2011}
\bibinfo{author}{\bibfnamefont{F.}~\bibnamefont{Wolfgramm}},
  \bibinfo{author}{\bibfnamefont{Y.~A.} \bibnamefont{de~Icaza~Astiz}},
  \bibinfo{author}{\bibfnamefont{F.~A.} \bibnamefont{Beduini}},
  \bibinfo{author}{\bibfnamefont{A.}~\bibnamefont{Cer\`e}}, \bibnamefont{and}
  \bibinfo{author}{\bibfnamefont{M.~W.} \bibnamefont{Mitchell}},
  \bibinfo{journal}{Phys. Rev. Lett.} \textbf{\bibinfo{volume}{106}},
  \bibinfo{pages}{053602} (\bibinfo{year}{2011}).

\bibitem[{\citenamefont{White et~al.}(1998)\citenamefont{White, Mitchell,
  Nairz, and Kwiat}}]{Kwiat:1998}
\bibinfo{author}{\bibfnamefont{A.~G.} \bibnamefont{White}},
  \bibinfo{author}{\bibfnamefont{J.~R.} \bibnamefont{Mitchell}},
  \bibinfo{author}{\bibfnamefont{O.}~\bibnamefont{Nairz}}, \bibnamefont{and}
  \bibinfo{author}{\bibfnamefont{P.~G.} \bibnamefont{Kwiat}},
  \bibinfo{journal}{Phys. Rev. A} \textbf{\bibinfo{volume}{58}},
  \bibinfo{pages}{605} (\bibinfo{year}{1998}).

\bibitem[{\citenamefont{Vaidman}(2007)}]{Vaidman:2007}
\bibinfo{author}{\bibfnamefont{L.}~\bibnamefont{Vaidman}},
  \bibinfo{journal}{Phys. Rev. Lett.} \textbf{\bibinfo{volume}{98}},
  \bibinfo{pages}{160403} (\bibinfo{year}{2007}).

\bibitem[{\citenamefont{Paraoanu}(2006)}]{Paraoanu:2006}
\bibinfo{author}{\bibfnamefont{G.~S.} \bibnamefont{Paraoanu}},
  \bibinfo{journal}{Phys. Rev. Lett.} \textbf{\bibinfo{volume}{97}},
  \bibinfo{pages}{180406} (\bibinfo{year}{2006}).

\bibitem[{\citenamefont{Strambini et~al.}(2010)\citenamefont{Strambini,
  Chirolli, Giovannetti, Taddei, Fazio, Piazza, and Beltram}}]{Strambini:2010}
\bibinfo{author}{\bibfnamefont{E.}~\bibnamefont{Strambini}},
  \bibinfo{author}{\bibfnamefont{L.}~\bibnamefont{Chirolli}},
  \bibinfo{author}{\bibfnamefont{V.}~\bibnamefont{Giovannetti}},
  \bibinfo{author}{\bibfnamefont{F.}~\bibnamefont{Taddei}},
  \bibinfo{author}{\bibfnamefont{R.}~\bibnamefont{Fazio}},
  \bibinfo{author}{\bibfnamefont{V.}~\bibnamefont{Piazza}}, \bibnamefont{and}
  \bibinfo{author}{\bibfnamefont{F.}~\bibnamefont{Beltram}},
  \bibinfo{journal}{Phys. Rev. Lett.} \textbf{\bibinfo{volume}{104}},
  \bibinfo{pages}{170403} (\bibinfo{year}{2010}).

\bibitem[{\citenamefont{Chirolli et~al.}(2010)\citenamefont{Chirolli,
  Strambini, Giovannetti, Taddei, Piazza, Fazio, Beltram, and
  Burkard}}]{Chirolli:2010}
\bibinfo{author}{\bibfnamefont{L.}~\bibnamefont{Chirolli}},
  \bibinfo{author}{\bibfnamefont{E.}~\bibnamefont{Strambini}},
  \bibinfo{author}{\bibfnamefont{V.}~\bibnamefont{Giovannetti}},
  \bibinfo{author}{\bibfnamefont{F.}~\bibnamefont{Taddei}},
  \bibinfo{author}{\bibfnamefont{V.}~\bibnamefont{Piazza}},
  \bibinfo{author}{\bibfnamefont{R.}~\bibnamefont{Fazio}},
  \bibinfo{author}{\bibfnamefont{F.}~\bibnamefont{Beltram}}, \bibnamefont{and}
  \bibinfo{author}{\bibfnamefont{G.}~\bibnamefont{Burkard}},
  \bibinfo{journal}{Phys. Rev. B} \textbf{\bibinfo{volume}{82}},
  \bibinfo{pages}{045403} (\bibinfo{year}{2010}).

\bibitem[{\citenamefont{Halperin}(1982)}]{Halperin1982}
\bibinfo{author}{\bibfnamefont{B.~I.} \bibnamefont{Halperin}},
  \bibinfo{journal}{Phys. Rev. B} \textbf{\bibinfo{volume}{25}},
  \bibinfo{pages}{2185} (\bibinfo{year}{1982}), \eprint{9506066v2}.

\bibitem[{\citenamefont{Wen}(1990)}]{Wen1990}
\bibinfo{author}{\bibfnamefont{X.~G.} \bibnamefont{Wen}},
  \bibinfo{journal}{Phys. Rev. B} \textbf{\bibinfo{volume}{41}},
  \bibinfo{pages}{12838} (\bibinfo{year}{1990}).

\bibitem[{\citenamefont{von Delft and Schoeller}(1998)}]{vonDelft:1998}
\bibinfo{author}{\bibfnamefont{J.}~\bibnamefont{von Delft}} \bibnamefont{and}
  \bibinfo{author}{\bibfnamefont{H.}~\bibnamefont{Schoeller}},
  \bibinfo{journal}{Ann. Phys.} \textbf{\bibinfo{volume}{7}},
  \bibinfo{pages}{225} (\bibinfo{year}{1998}).

\bibitem[{\citenamefont{Texier and B\"uttiker}(2000)}]{PhysRevB.62.7454}
\bibinfo{author}{\bibfnamefont{C.}~\bibnamefont{Texier}} \bibnamefont{and}
  \bibinfo{author}{\bibfnamefont{M.}~\bibnamefont{B\"uttiker}},
  \bibinfo{journal}{Phys. Rev. B} \textbf{\bibinfo{volume}{62}},
  \bibinfo{pages}{7454} (\bibinfo{year}{2000}).

\bibitem[{\citenamefont{Ji et~al.}(2003)\citenamefont{Ji, Chung, Sprinzak,
  Heiblum, Mahalu, and Shtrikman}}]{ji2003electronic}
\bibinfo{author}{\bibfnamefont{Y.}~\bibnamefont{Ji}},
  \bibinfo{author}{\bibfnamefont{Y.}~\bibnamefont{Chung}},
  \bibinfo{author}{\bibfnamefont{D.}~\bibnamefont{Sprinzak}},
  \bibinfo{author}{\bibfnamefont{M.}~\bibnamefont{Heiblum}},
  \bibinfo{author}{\bibfnamefont{D.}~\bibnamefont{Mahalu}}, \bibnamefont{and}
  \bibinfo{author}{\bibfnamefont{H.}~\bibnamefont{Shtrikman}},
  \bibinfo{journal}{Nature} \textbf{\bibinfo{volume}{422}},
  \bibinfo{pages}{415} (\bibinfo{year}{2003}).

\bibitem[{\citenamefont{Dubois et~al.}(2013)\citenamefont{Dubois, Jullien,
  Portier, Roche, Cavanna, Jin, Wegscheider, Roulleau, and
  Glattli}}]{Dubois:2013}
\bibinfo{author}{\bibfnamefont{J.}~\bibnamefont{Dubois}},
  \bibinfo{author}{\bibfnamefont{T.}~\bibnamefont{Jullien}},
  \bibinfo{author}{\bibfnamefont{F.}~\bibnamefont{Portier}},
  \bibinfo{author}{\bibfnamefont{P.}~\bibnamefont{Roche}},
  \bibinfo{author}{\bibfnamefont{A.}~\bibnamefont{Cavanna}},
  \bibinfo{author}{\bibfnamefont{Y.}~\bibnamefont{Jin}},
  \bibinfo{author}{\bibfnamefont{W.}~\bibnamefont{Wegscheider}},
  \bibinfo{author}{\bibfnamefont{P.}~\bibnamefont{Roulleau}}, \bibnamefont{and}
  \bibinfo{author}{\bibfnamefont{D.}~\bibnamefont{Glattli}},
  \bibinfo{journal}{Nature} \textbf{\bibinfo{volume}{502}},
  \bibinfo{pages}{659} (\bibinfo{year}{2013}).

\bibitem[{\citenamefont{Mitchison and Massar}(2001)}]{Mitchison:2001}
\bibinfo{author}{\bibfnamefont{G.}~\bibnamefont{Mitchison}} \bibnamefont{and}
  \bibinfo{author}{\bibfnamefont{S.}~\bibnamefont{Massar}},
  \bibinfo{journal}{Phys. Rev. A} \textbf{\bibinfo{volume}{63}},
  \bibinfo{pages}{032105} (\bibinfo{year}{2001}).

\bibitem[{Not()}]{Note1}
\bibinfo{note}{We assume here and throughout the manuscript that the scattering
  matrix elements are independent on $k$.}

\bibitem[{\citenamefont{Pryadko and Korotkov}(2007)}]{Korotkov:2007}
\bibinfo{author}{\bibfnamefont{L.~P.} \bibnamefont{Pryadko}} \bibnamefont{and}
  \bibinfo{author}{\bibfnamefont{A.~N.} \bibnamefont{Korotkov}},
  \bibinfo{journal}{Phys. Rev. B} \textbf{\bibinfo{volume}{76}},
  \bibinfo{pages}{100503} (\bibinfo{year}{2007}).

\bibitem[{\citenamefont{Katz et~al.}(2008)\citenamefont{Katz, Neeley, Ansmann,
  Bialczak, Hofheinz, Lucero, O'Connell, Wang, Cleland, Martinis
  et~al.}}]{Katz08}
\bibinfo{author}{\bibfnamefont{N.}~\bibnamefont{Katz}},
  \bibinfo{author}{\bibfnamefont{M.}~\bibnamefont{Neeley}},
  \bibinfo{author}{\bibfnamefont{M.}~\bibnamefont{Ansmann}},
  \bibinfo{author}{\bibfnamefont{R.~C.} \bibnamefont{Bialczak}},
  \bibinfo{author}{\bibfnamefont{M.}~\bibnamefont{Hofheinz}},
  \bibinfo{author}{\bibfnamefont{E.}~\bibnamefont{Lucero}},
  \bibinfo{author}{\bibfnamefont{A.}~\bibnamefont{O'Connell}},
  \bibinfo{author}{\bibfnamefont{H.}~\bibnamefont{Wang}},
  \bibinfo{author}{\bibfnamefont{A.~N.} \bibnamefont{Cleland}},
  \bibinfo{author}{\bibfnamefont{J.~M.} \bibnamefont{Martinis}},
  \bibnamefont{et~al.}, \bibinfo{journal}{Phys. Rev. Lett.}
  \textbf{\bibinfo{volume}{101}}, \bibinfo{pages}{200401}
  (\bibinfo{year}{2008}).

\bibitem[{\citenamefont{Zilberberg et~al.}(2013)\citenamefont{Zilberberg,
  Romito, Starling, Howland, Broadbent, Howell, and Gefen}}]{Zilberberg:2013}
\bibinfo{author}{\bibfnamefont{O.}~\bibnamefont{Zilberberg}},
  \bibinfo{author}{\bibfnamefont{A.}~\bibnamefont{Romito}},
  \bibinfo{author}{\bibfnamefont{D.~J.} \bibnamefont{Starling}},
  \bibinfo{author}{\bibfnamefont{G.~A.} \bibnamefont{Howland}},
  \bibinfo{author}{\bibfnamefont{C.~J.} \bibnamefont{Broadbent}},
  \bibinfo{author}{\bibfnamefont{J.~C.} \bibnamefont{Howell}},
  \bibnamefont{and} \bibinfo{author}{\bibfnamefont{Y.}~\bibnamefont{Gefen}},
  \bibinfo{journal}{Physical review letters} \textbf{\bibinfo{volume}{110}},
  \bibinfo{pages}{170405} (\bibinfo{year}{2013}).

\bibitem[{\citenamefont{Zilberberg et~al.}(2012)\citenamefont{Zilberberg,
  Romito, and Gefen}}]{zilberbergCrypta}
\bibinfo{author}{\bibfnamefont{O.}~\bibnamefont{Zilberberg}},
  \bibinfo{author}{\bibfnamefont{A.}~\bibnamefont{Romito}}, \bibnamefont{and}
  \bibinfo{author}{\bibfnamefont{Y.}~\bibnamefont{Gefen}},
  \bibinfo{journal}{Physica Scripta} \textbf{\bibinfo{volume}{2012}},
  \bibinfo{pages}{014014} (\bibinfo{year}{2012}).

\bibitem[{\citenamefont{Zilberberg et~al.}(2014)\citenamefont{Zilberberg,
  Romito, and Gefen}}]{zilberberg2014standard}
\bibinfo{author}{\bibfnamefont{O.}~\bibnamefont{Zilberberg}},
  \bibinfo{author}{\bibfnamefont{A.}~\bibnamefont{Romito}}, \bibnamefont{and}
  \bibinfo{author}{\bibfnamefont{Y.}~\bibnamefont{Gefen}}, in
  \emph{\bibinfo{booktitle}{Quantum Theory: A Two-Time Success Story}}
  (\bibinfo{publisher}{Springer}, \bibinfo{year}{2014}), pp.
  \bibinfo{pages}{377--387}.

\bibitem[{\citenamefont{Neder et~al.}(2007)\citenamefont{Neder, Ofek, Chung,
  Heiblum, Mahalu, and Umansky}}]{Neder:2007}
\bibinfo{author}{\bibfnamefont{I.}~\bibnamefont{Neder}},
  \bibinfo{author}{\bibfnamefont{N.}~\bibnamefont{Ofek}},
  \bibinfo{author}{\bibfnamefont{Y.}~\bibnamefont{Chung}},
  \bibinfo{author}{\bibfnamefont{M.}~\bibnamefont{Heiblum}},
  \bibinfo{author}{\bibfnamefont{D.}~\bibnamefont{Mahalu}}, \bibnamefont{and}
  \bibinfo{author}{\bibfnamefont{V.}~\bibnamefont{Umansky}},
  \bibinfo{journal}{Nature} \textbf{\bibinfo{volume}{448}},
  \bibinfo{pages}{333} (\bibinfo{year}{2007}).

\bibitem[{\citenamefont{Blanter and Buttiker}(2000)}]{Blanter:2000}
\bibinfo{author}{\bibfnamefont{Y.~M.} \bibnamefont{Blanter}} \bibnamefont{and}
  \bibinfo{author}{\bibfnamefont{M.}~\bibnamefont{Buttiker}},
  \bibinfo{journal}{Phys. Rep.} \textbf{\bibinfo{volume}{336}},
  \bibinfo{pages}{1} (\bibinfo{year}{2000}).

\end{thebibliography}


\end{document}